\documentclass{emulateapj}
\usepackage{multirow,natbib,epsf}

\newcommand{\Msun}{$M_{\sun}$}
\newcommand{\Lsun}{$L_{\sun}$}

\newcommand{\Ha}{H$\alpha$\ }

\newcommand{\sfg}{S$^{4}$G\ }
\newcommand{\sfgn}{S$^{4}$G}

\slugcomment{}
\shorttitle{The Spitzer Survey of Stellar Structure in Galaxies
  (S$^4$G)}
\shortauthors{Sheth et al.}

\begin{document}


\title{The Spitzer Survey of Stellar Structure in Galaxies
  (S$^4$G)}

\author{Kartik Sheth\altaffilmark{1,2,3},
Michael Regan\altaffilmark{4},   
Joannah L. Hinz\altaffilmark{5},
Armando Gil de Paz\altaffilmark{6},
Kar\'{i}n Men\'{e}ndez-Delmestre\altaffilmark{7},
Juan-Carlos Mu\~noz-Mateos\altaffilmark{1,6},
Mark Seibert\altaffilmark{7},
Taehyun Kim\altaffilmark{1,24}
Eija Laurikainen\altaffilmark{8,9},  
Heikki Salo\altaffilmark{8},  
Dimitri A. Gadotti\altaffilmark{10,11},
Jarkko Laine\altaffilmark{8,9},
Trisha Mizusawa\altaffilmark{1,2,3},
Lee Armus\altaffilmark{2,3},
E. Athanassoula\altaffilmark{12},
Albert Bosma\altaffilmark{12},
Ronald J. Buta\altaffilmark{13},
Peter Capak\altaffilmark{2,3},
Thomas H. Jarrett\altaffilmark{2,3},
Debra M. Elmegreen\altaffilmark{14},
Bruce G. Elmegreen\altaffilmark{15},  
Johan H. Knapen\altaffilmark{16,24},
Jin Koda\altaffilmark{17},
George Helou\altaffilmark{2,3},
Luis C. Ho\altaffilmark{7},
Barry F. Madore\altaffilmark{7},  
Karen L. Masters\altaffilmark{18},
Bahram Mobasher\altaffilmark{19},
Patrick Ogle\altaffilmark{2,3},
Chien Y. Peng\altaffilmark{20},
Eva Schinnerer\altaffilmark{21},  
Jason A. Surace\altaffilmark{2,3},
Dennis Zaritsky\altaffilmark{5},
S\'ebastien Comer\'on\altaffilmark{16,22},
Bonita de Swardt\altaffilmark{23}
Sharon E. Meidt\altaffilmark{21},
Mansi Kasliwal\altaffilmark{3},
Manuel Aravena\altaffilmark{1}
}

\altaffiltext{1}{National Radio Astronomy Observatory / NAASC, 520 Edgemont Road, Charlottesville, VA 22903}
\altaffiltext{2}{Spitzer Science Center}
\altaffiltext{3}{California Institute of Technology, 1200 East California Boulevard, Pasadena, CA 91125}
\altaffiltext{4}{Space Telescope Science Institute, 3700 San Martin Drive, Baltimore, MD 21218}
\altaffiltext{5}{University of Arizona, 933 N. Cherry Ave, Tucson, AZ  85721}
\altaffiltext{6}{Departamento de Astrof'sica, Universidad Complutense de Madrid, Madrid 28040, Spain}
\altaffiltext{7}{The Observatories, Carnegie Institution of Washington, 813 Santa Barbara Street, Pasadena, CA 91101}
\altaffiltext{8}{Department of Physical Sciences/Astronomy Division, University of Oulu, FIN-90014, Finland}
\altaffiltext{9}{Finnish Centre for Astronomy with ESO (FINCA), University of Turku}
\altaffiltext{10}{Max-Planck-Institut f\"ur Astrophysik, Karl-Schwarzschild-Strasse 1, D-85748 Garching bei MŸnchen, Germany}
\altaffiltext{11}{European Southern Observatory, Casilla 19001, Santiago 19, Chile}
\altaffiltext{12}{Laboratoire d'Astrophysique de Marseille (LAM), UMR6110, Universit\'e de Provence/CNRS, Technop\^ole de Marseille Etoile, 38 rue Fr\'ed\'eric Joliot Curie, 13388 Marseille C\'edex 20, France}
\altaffiltext{13}{Department of Physics and Astronomy, University of Alabama, Box 870324, Tuscaloosa, AL 35487, USA}
\altaffiltext{14}{Vassar College, Dept. of Physics and Astronomy, Poughkeepsie, NY 12604}
\altaffiltext{15}{IBM Research Division, T.J. Watson Research Center, Yorktown Hts., NY 10598}
\altaffiltext{16}{Departamento de Astrof\'\i sica, Universidad de La Laguna, Spain}
\altaffiltext{17}{Department of Physics and Astronomy, SUNY Stony Brook, Stony Brook, NY 11794-3800}
\altaffiltext{18}{Institute of Cosmology and Gravitation, University of Portsmouth, Dennis Sciama Building, Burnaby Road, Portsmouth, PO1 2EH, UK}
\altaffiltext{19}{Department of Physics and Astronomy, University of California, Riverside, CA 92521}
\altaffiltext{20}{NRC Herzberg Institute of Astrophysics, 5071 West Saanich Road, Victoria, V9E 2E7, Canada}
\altaffiltext{21}{Max-Planck-Institut f\"ur Astronomie, K\"onigstuhl 17, 69117 Heidelberg, Germany}
\altaffiltext{22}{Korea Astronomy and Space Science Institute, 838, Daedeokdae-ro, Yuseong-gu, Daejeon, 305-348, Republic of \altaffiltext{24}{Astronomy Program, Department of Physics and Astronomy, Seoul National University, Seoul 151-742, Korea}
\altaffiltext{23}{South African Astronomical Observatory, Observatory, 7935 Cape Town, South Africa}
\altaffiltext{24}{Instituto de Astrof\'\i sica de Canarias, E-38200 La Laguna, Spain}
 Korea}
\begin{abstract}
  The {\em Spitzer} Survey of Stellar Structure in Galaxies (S$^{4}$G) is an
  Exploration Science Legacy Program approved for the {\em Spitzer}
  post-cryogenic mission.  It is a volume-, magnitude-, and
  size-limited (d$<$40 Mpc, $|$b$|$ $>$ 30$^{\circ}$, m$_{Bcorr}<$15.5,
  D$_{25}>$1$\arcmin$) survey of 2,331 galaxies using the Infrared Array Camera (IRAC) at 3.6 and 4.5$\mu$m.  Each galaxy is observed for 240 s and
  mapped to $\ge$ 1.5$\times$D$_{25}$.  The final mosaicked images have a typical 1$\sigma$ rms noise level of 0.0072 and 0.0093 MJy sr$^{-1}$ at 3.6 and 4.5$\mu$m, respectively.  Our azimuthally-averaged surface brightness profile typically traces isophotes at $\mu_{3.6\mu m}(AB) (1 \sigma) \sim$ 27 mag arcsec$^{-2}$, equivalent to a stellar mass surface density of $\sim$ 1 \Msun pc$^{-2}$.     \sfg thus provides an unprecedented
  data set for the study of the distribution of mass and stellar structures in the local Universe.  
  This large, unbiased and extremely deep sample of all Hubble types from dwarfs to spirals
  to ellipticals will allow for detailed structural studies, not only as a
  function of stellar mass, but also as a function of the local
  environment. The data from this survey will serve as a vital
  testbed for cosmological simulations predicting the stellar mass properties
  of present-day galaxies.  This paper introduces the survey,
  describes the sample selection, the significance of the 3.6 and
  4.5$\mu$m bands for this study, and the data collection \& 
  survey strategy.  
  We describe the \sfg data analysis pipeline and present measurements for a first set of galaxies, observed in both the cryogenic and warm mission phase of {\em Spitzer}. For every galaxy we tabulate the galaxy diameter, position angle, axial ratio, inclination at $\mu_{3.6\mu m} (AB)$= 25.5 and 26.5 mag arcsec$^{-2}$ (equivalent to $\approx \mu_\mathrm{B} (AB)$=27.2 and 28.2 mag arcsec$^{-2}$, respectively). These measurements will form the initial \sfg catalog of galaxy properties.  We also measure the total magnitude and the azimuthally-averaged radial profiles of ellipticity, position angle, surface brightness and color.  Finally, we deconstruct each galaxy using GALFIT  into its main constituent stellar components: the  bulge/spheroid, disk,  bar, and nuclear point source, where necessary.   Together these data products will provide  a comprehensive and definitive catalog of stellar structures, mass and properties of galaxies in the nearby Universe and enable a variety of scientific investigations, some of which are highlighted in this introductory \sfg survey paper.

\end{abstract}

\section{Introduction}

Understanding the distribution of stars within a galaxy is akin to the study of the endoskeleton of a body; embedded within the galaxy is the fossil record of the assembly history and evolutionary processes of cosmic  time.  The first step in unraveling this history is to obtain a complete census of the stellar structures in galaxies in the local volume.  This is the primary motivation for the {\em Spitzer} Survey of Stellar Structures in Galaxies (\sfgn).  \sfg is a volume-limited (d $<$ 40 Mpc), size-limited (D$_{25} > $1$\arcmin$) and apparent B-band brightness (corrected for inclination, galactic extinction and
  K-correction) limited (m$_{Bcorr} < $15.5) survey of 2,331 nearby galaxies at 3.6 and 4.5$\mu$m with the Infrared Array Camera (IRAC) \citep{fazio04} on the {\em Spitzer} Space Telescope (SSC) \citep{werner04}.
 
   

Over the last 70 years, numerous surveys have sought to create
baseline data sets for nearby galaxies with as few as tens of galaxies (ANGST: \citealt{dalcanton09}; SINGS, \citealt{kennicutt03}) to tens of thousands of galaxies (e.g., RC3, \citealt{devaucouleurs91}). While these surveys was designed specifically for particular scientific goals, none of them provides an accurate inventory of the stellar mass and the stellar structures in nearby galaxies. The main reason is that infrared surveys, where light from the old stellar population is better  measured due to reduced dust extinction and contamination from star formation, are extremely difficult to conduct from the ground. The largest ground-based infrared surveys (e.g., 2MASS, \citealt{skrutskie06}; DENIS, \citealt{epchtein94,paturel03}) are rather shallow in depth, and as a result, studies of nearby galaxies have been restricted to high surface brightness inner disks.  In contrast, deeper infrared surveys (e.g., OSUBGS, \citealt{eskridge02}, NIRSOS, \citealt{laurikainen10}) have imaged only a few hundred galaxies.  \sfg is specifically designed to provide over an order of magnitude improvement in the sample size and several magnitudes deeper data than existing surveys.  

\begin{figure}
\plottwo{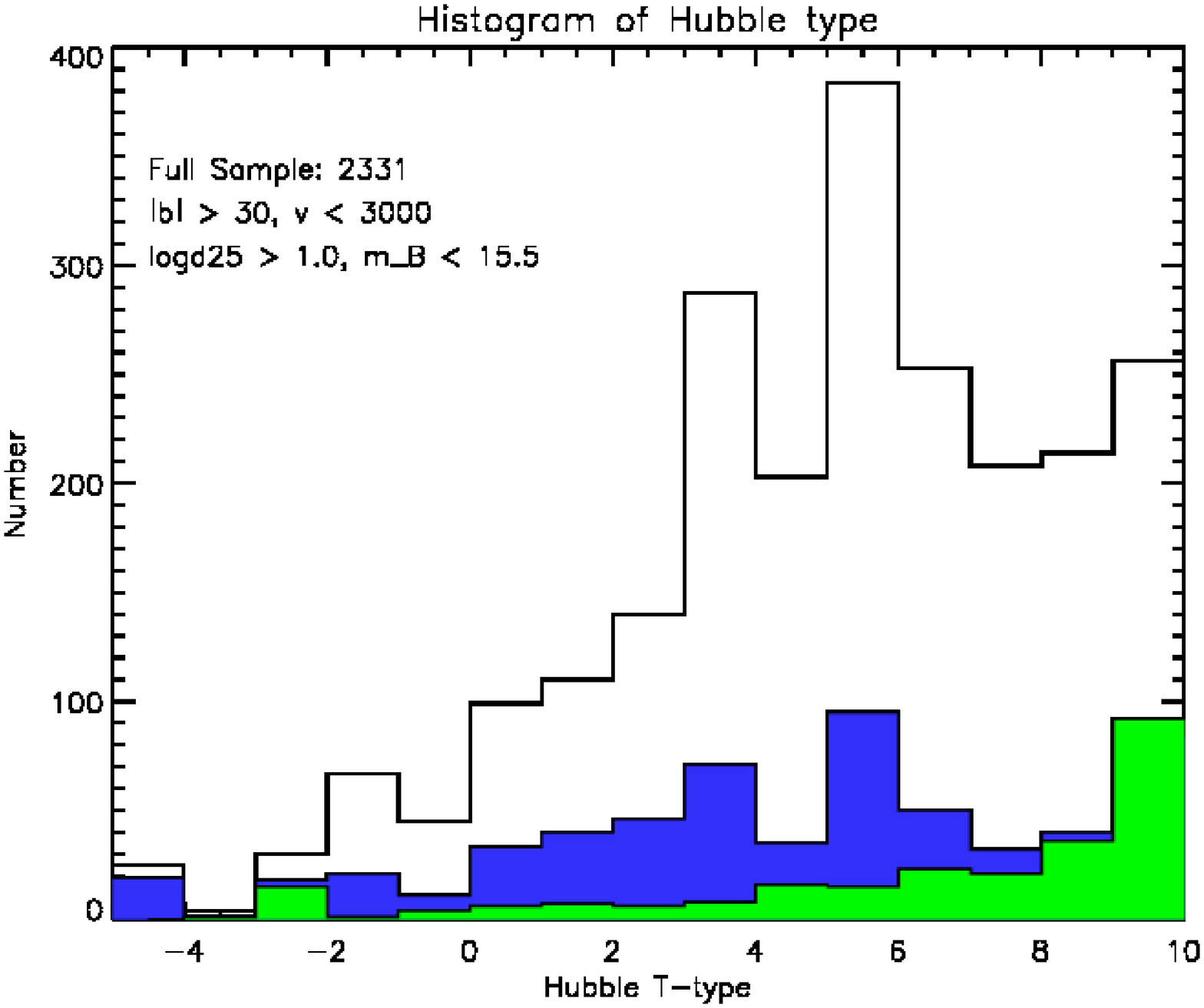}{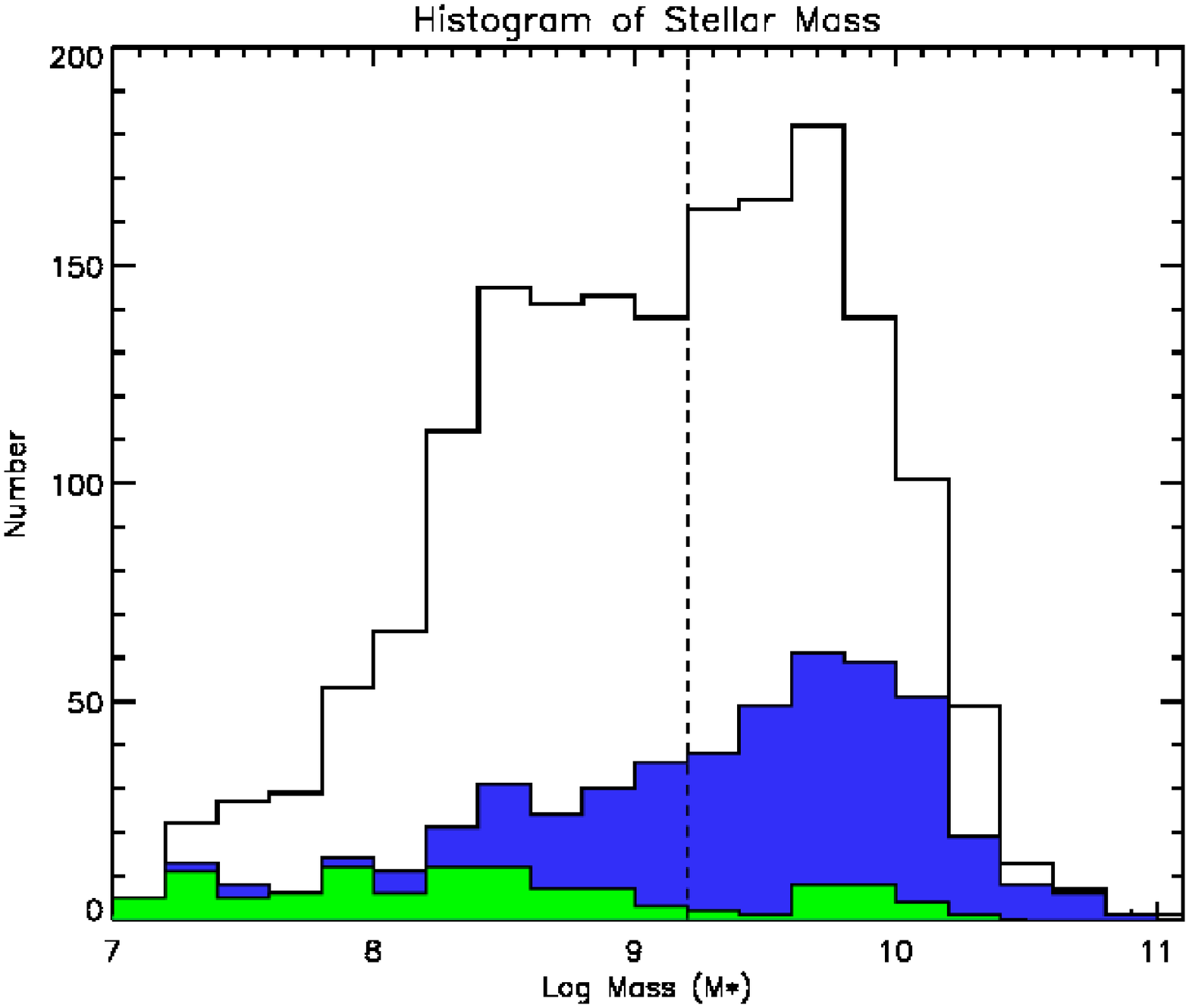}
\caption{ {\em Left:} Histogram of all galaxies in \sfg as a function of the Hubble
  T-type shown in black. The green and blue filled
  histograms show archival data from the LVL (green) and other GO/GTO
  data (blue). {\em Right:} Histogram of masses for the \sfg sample galaxies shown
  in black, where masses are calculated from the 2MASS photometry assuming the M/L function from \citep{bell03}.  The biggest gains in building a statistically complete sample are for galaxies with log (M/\Msun) $<$ 9.2.}\label{hubmasstyp}
\end{figure}

\begin{figure}
\plotone{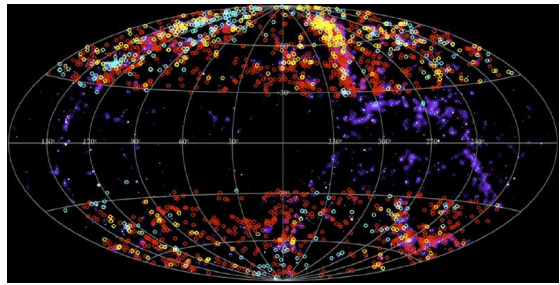}
\caption{Galaxies from \sfg plotted on the local large scale
  structure map from the 2MASS XSC \citep{jarrett04} shown in purple. The red circles show the warm
  mission \sfg galaxies to be observed. The cyan and yellow circles
  show LVL and GO/GTO archival galaxies. With the full \sfg sample, we will have the ability to make a less biased study of the galaxy host properties as a function of the local large scale structures. } \label{volume}
\end{figure}

\sfg builds upon the two previous
{\em Spitzer} Legacy surveys of nearby galaxies: SINGS \citep{kennicutt03}
and the Local Volume Legacy Survey (LVL) \citep{lee08}. SINGS was designed to probe
star formation, dust and polycyclic aromatic hydrocarbon (PAH) emission in all representative
environments {\em within} galaxies in a sample of 75 objects. The LVL
survey was designed to study spatially-resolved star formation and the
red stellar population within galaxies in a local volume of 11
Mpc. The local volume limits the 258 LVL galaxies to consist mostly of
dwarf, irregular and late type systems (see green histogram in
Fig.\ 1). The remainder of the {\em Spitzer} archive from the cryogenic mission is
inadequate for the science goals outlined here because it lacks
sufficient numbers of galaxies of all stellar masses, particularly  
at M$_{*} < $ 10$^9$ \Msun, Hubble types and environments. As shown in 
Figures \ref{hubmasstyp} and \ref{volume}, \sfg explores the full range of stellar structures
in a representative and large sample of galaxies of all types, masses and in diverse
environments.  \sfg is designed to provide the ultimate baseline data set for the study of stellar structures and mass in nearby galaxies.

In this paper we introduce \sfgn, describe the data collection and analysis strategy and briefly describe the variety of scientific investigations possible with these data.  All \sfg galaxies are processed through a uniform pipeline to create the deepest and largest mid-infrared image catalog of nearby galaxies to date.  For every galaxy we measure the magnitude and surface brightness and tabulate the galaxy diameter, position angle, axial ratio, and inclination at $\mu_{3.6\mu m}$ (AB) = 25.5 and 26.5 mag arcsec$^{-2}$.  We also compute the azimuthally-averaged radial profiles of ellipticity, position angle, surface brightness and color.  This fundamental catalog of galaxy properties will allow us to classify
their  morphology based on their stellar structures -- a preliminary study is presented in \citet{buta10}.  
In the long term, the main goal of the \sfg project is to obtain a  detailed understanding of the properties of the different stellar structures, their formation and  evolutionary paths and their role in the broader picture of galaxy evolution.  This will be done by decomposing every galaxy into its constituent structural components (e.g., bulge/spheroid, bar, disk, nuclear point source) using GALFIT \citep{peng02,peng10}.  

A number of scientific studies are already under way by the \sfg team.  These include a study of outer disks to determine the variety and frequency of different disk profile(s), truncations or thresholds and their relationship to the build up of galaxy disks, a  study of shells and debris around galaxies and their relationship to central star formation or AGN activity, an analysis of the structural properties of bulges/spheroids (classical, disky, boxy/peanut-shaped), determination of the bar fraction and bar properties in relationship to the host galaxies and environment, an analysis of the stellar arm-interarm variations in spirals, an investigation of the underlying old stellar population in rings and a search for fossil rings, an analysis of the old stellar content of dwarf galaxies and its implication for their formation history, etc.  
Longer term studies include star formation history and assembly of galaxies using {\it GALEX} and \Ha data, the Tully--Fisher relationship using existing HI data (e.g., ALFALFA, \citealt{giovanelli05}, HIPASS, \citealt{barnes01}), and testing of new mid-infrared diagnostics of AGN.  We describe these studies in more detail in \S \ref{sfgsci}.  Ultimately, \sfg will lead to numerous other studies of astrophysical phenomena in the nearby Universe and provide a key anchoring data set for studies of galaxy evolution.

%

\section{The {\em Spitzer} 3.6 and 4.5$\mu$m Bands: Best Tracers of Stellar Mass}

The 3.6 and 4.5$\mu$m IRAC bands on {\em Spitzer} are ideal tracers of
the stellar mass distribution in galaxies because they image the
Rayleigh--Jeans limit of the blackbody emission for stars with T $>$
2000 K. Moreover, the mid-infrared [3.6]--[4.5] color in nearby
galaxies is nearly constant with radius, and independent of the age of the stellar
population and its mass function \citep{pahre04}.  Analysis of the tilt in the fundamental plane for ellipticals shows that the {\em Spitzer} IRAC bands are better tracers of the stellar mass than even K-band emission, which is more sensitive to variations in the underlying stellar population; the IRAC data are more uniform and therefore are the best and perhaps most direct tracers of stellar mass \citep {jun08}.  Additionally, on {\em Spitzer} these bands are so sensitive that \sfg will image to extremely low stellar mass surface densities ($\sim$ 1\Msun pc$^{-2}$). 

There are minor contaminants to the old stellar population light in the 3.6 and 4.5$\mu$m IRAC bands.  
In the 3.6$\mu$m band, there is a very weak 3.3$\mu$m PAH feature, which contributes negligibly to the overall emission in the band ($<$ 2\% as the equivalent width is $\sim$0.02$\mu$m, \citealt{tokunaga91}).  Hot dust (T$_{d} > $500K) from
very small grains can contribute light to both bands with a higher contribution in the 4.5$\mu$m band, but this condition occurs only near active galactic nuclei or extreme starbursts.  By examining the colors of the galaxies, we should be able to use the two IRAC bands to remove the 
effects of the 3.3$\mu$m PAH or hot dust emission.  Another important advantage of having 2 bands is that by co-adding them, we can further improve the signal to noise and push the study to even fainter regions of the galaxies.   In both bands, the contamination from young red supergiants is also very low.  From previous studies,  it is known that in the near-infrared, the overall
contamination from young red supergiants to a galaxy is only $\sim$3\%
but in regions of star formation, the contribution may be as high as 25\% \citep{rhoads98}.  These regions, however, are easily identifiable and can be flagged.  At the IRAC wavelengths, the fractional contribution from star-forming regions is expected to be the same (or lower) because these wavebands are  
at longer wavelengths on the Rayleigh--Jeans tail of the spectral energy distribution.  At these wavelengths,  there is lower dust extinction \citep{draine84} and no other significant emission sources. Thus, with the extremely low surface brightness limits, \sfg offers a unique and virtually dust-free view of the distribution of mass in stellar structure in the nearby Universe.

Although ground-based near-infrared observations can offer higher angular
resolution than {\em Spitzer}, the main obstacle for ground-based
observations is the very high and variable sky brightness (typically
$\mu_{K} \sim$ 13.4 mag arcsec$^{-2}$). The surface brightness level
in the IRAC data is over 10 magnitudes below the typical sky
brightness level. To get to this level from the ground, one would need to characterize the sky
brightness, the flat field and instrumental variations to better
than 0.0009\%. This is currently not possible with any existing (or planned)
near-infrared survey. Surveys like UKIDSS LAS or VISTA VHS are planning to  reach a depth of 18.4 (21.2 in AB) mag arcsec$^{-2}$ in the K$_s$ band, which translates to 
30 \Lsun pc$^{-2}$ (even with the brighter stellar emission at K$_s$
band), whereas \sfg reaches $\sim$ 2 \Lsun pc $^{-2}$.  It is the
stability of the background that is the biggest advantage for
space-based observations.   

\section{Sample Selection}

We chose all galaxies with radial velocity V$_{radio}$ $<$ 3000 km/s (corresponding to a distance d $<$ 40 Mpc for a Hubble constant of 75 km/s/Mpc), total corrected blue magnitude m$_{Bcorr} <$ 15.5, blue light isophotal angular diameter D$_{25}$
$>$ 1.0 arcmin at galactic latitude $|$b$|$ $>$ 30$^{\circ}$ using HyperLEDA \citep{paturel03}. The choice of V$_{radio}$ does limit the sample to galaxies with a radio-derived  (e.g., HI) radial velocities in HyperLEDA, and as a result misses some galaxies for which only optically-derived radial velocities exists; currently, a sample chosen with V$_{optical}$ instead of V$_{radio}$ would contain 2997 galaxies.  A comparison of the galaxy properties of the two samples shows that the \sfg sample misses galaxies that are small, relatively faint and early-type (gas-poor) systems from the volume surveyed.  

We also note that our sample was defined using HyperLEDA in September 2007 - since then 98 more galaxies have been added to the HyperLEDA data base that meet the \sfg criteria described above and we expect that some more may be added as better data become available.  Given the limited lifetime of {\em Spitzer}, we are unable to go back and acquire data on these  additional galaxies but we expect that the few percent additional galaxies will not strongly influence the core characteristics of the sample, except for the bias against early-type galaxies as noted above.

The choice of a 40 Mpc volume is arbitrary -- it was chosen to be large enough to provide a statistically significant number of galaxies of all types and to be representative of a large range of
the local large scale structure environment (as shown in Fig. \ref{volume}).  
Our experience with previous surveys of galaxy properties (e.g., \citealt{sheth08}) has shown that
a few thousand galaxies are needed to accurately measure and
account for completeness effects in mass, color and size selection of galaxy samples for robust investigations of relationships between galactic structure and host galaxy properties. The size cut, log D$_{25} >$ 1.0, was made to ensure that galaxies were large enough for a detailed study of their internal structure (at 40 Mpc, 1$\arcmin \sim$11.6 kpc). The apparent size and apparent magnitude cut were chosen to match the RC3 limits. The galactic latitude cut, $|$b$|$ $>$ 30$^{\circ}$, was used to minimize the unresolved Galactic light contribution from the Milky Way disk. The full sample of 2,331 \sfg galaxies is listed in Table \ref{s4gsamptab}.  In Figure \ref{hubmasstyp}, we show the distribution of \sfg galaxies as a function of Hubble type and stellar mass, along with the existing distribution of the LVL galaxies (green) and GO/GTO galaxies (blue) including the SINGS galaxies.  There are some noteworthy sub-samples within \sfg.  There are 188 early-type galaxies (ellipticals and lenticulars), 206 dwarf galaxies (defined as M$_B > $ -17), of which 135 are DDO dwarfs and 465 edge-on ($i > $ 75$^{\circ}$) systems. Also over 800 of the \sfg galaxies are mapped beyond a diameter of 3$\times$D$_{25}$.  These sub-samples provide the deepest, largest and most homogeneous samples for a multitude of galaxy evolutionary studies, some of which are described in \S \ref{sfgsci}.


\subsection{Archival / Cryogenic {\em Spitzer} Data}

Of the 2,331 galaxies in the \sfg sample, 597 had some existing data
at 3.6 and 4.5$\mu$m from the {\em Spitzer} cryogenic mission.  125 of these
are part of the LVL survey and 56 are from the SINGS survey, both of which
used the same observing strategy (\S \ref{obsstrat}) we employ for \sfg.  Almost all of
the archival maps have at least 240 s of integration time per pixel and
are sufficiently deep.  The only exceptions are NGC 5457 (96s), NGC
0470, NGC 0474 (150s), and NGC 5218, NGC 5216, and NGC 5576 (192s).  

In terms of area, 82 of the archival galaxies were mapped between 1.0 and 1.5$\times$D$_{25}$, and 43 others were mapped to $<$1.0 $\times$D$_{25}$.  While outer disk science for these 125 galaxies would benefit from an extended map, they represent only a small fraction of the total sample, and the repetition
of observations would have yielded only an incremental  scientific return. Therefore, we decided not to repeat observations of any of the archival galaxies.  We also decided not to map the SMC and the LMC, which meet our selection criteria but are very large and are being observed as part of other GO programs.  M33, the other commonly studied Local Group galaxy was observed during the cryogenic mission and is included in the archival part of our sample.  All other galaxies that were found via the HyperLEDA data base search in 2007 September are observed (see note earlier about the modest additions to the HyperLEDA database in the last three years).  Although we do not repeat observations of any of the  archival galaxies, we are processing all galaxies in the \sfg sample, except the SMC and LMC, through the same \sfg reduction and analysis pipelines described in \S \ref{pipeline}. 

\section{{\em Spitzer} Post-Cryogenic Mission Observations Strategy} \label{obsstrat}

The 1,734 \sfg galaxies being observed in the post-cryogenic mission are mapped using either a small dithered map or a mosaicked observation with both IRAC channels.  All galaxies are observed with a total
on-source integration time of 240 s which leads to  an rms noise of $\mu$(1$\sigma$) $\sim$ 0.0072 and 0.0093 MJy sr$^{-1}$ at 3.6 and 4.5$\mu$m, respectively. This translates
to a typical surface brightness sensitivity of $\mu_{AB} \sim$27 mag arcsec$^{-2}$ or a stellar surface density of 1.13 \Msun pc$^{-2}$ (assuming a solar M$_{K}$ = 3.33, \citealt{worthey94};  M/L$_{K}$=1, \citealt{heraudeau97}) .  

\begin{figure}
\plottwo{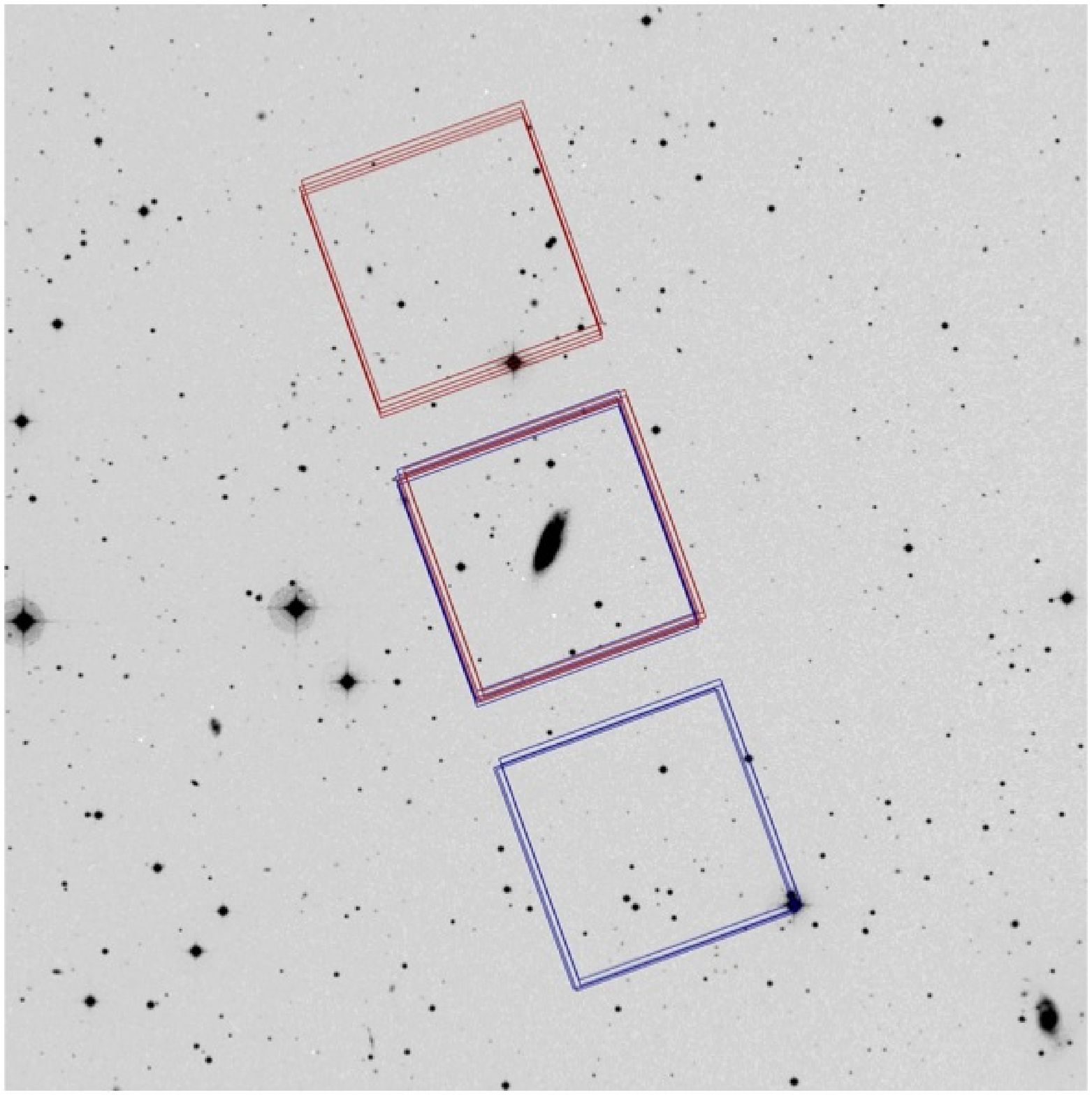}{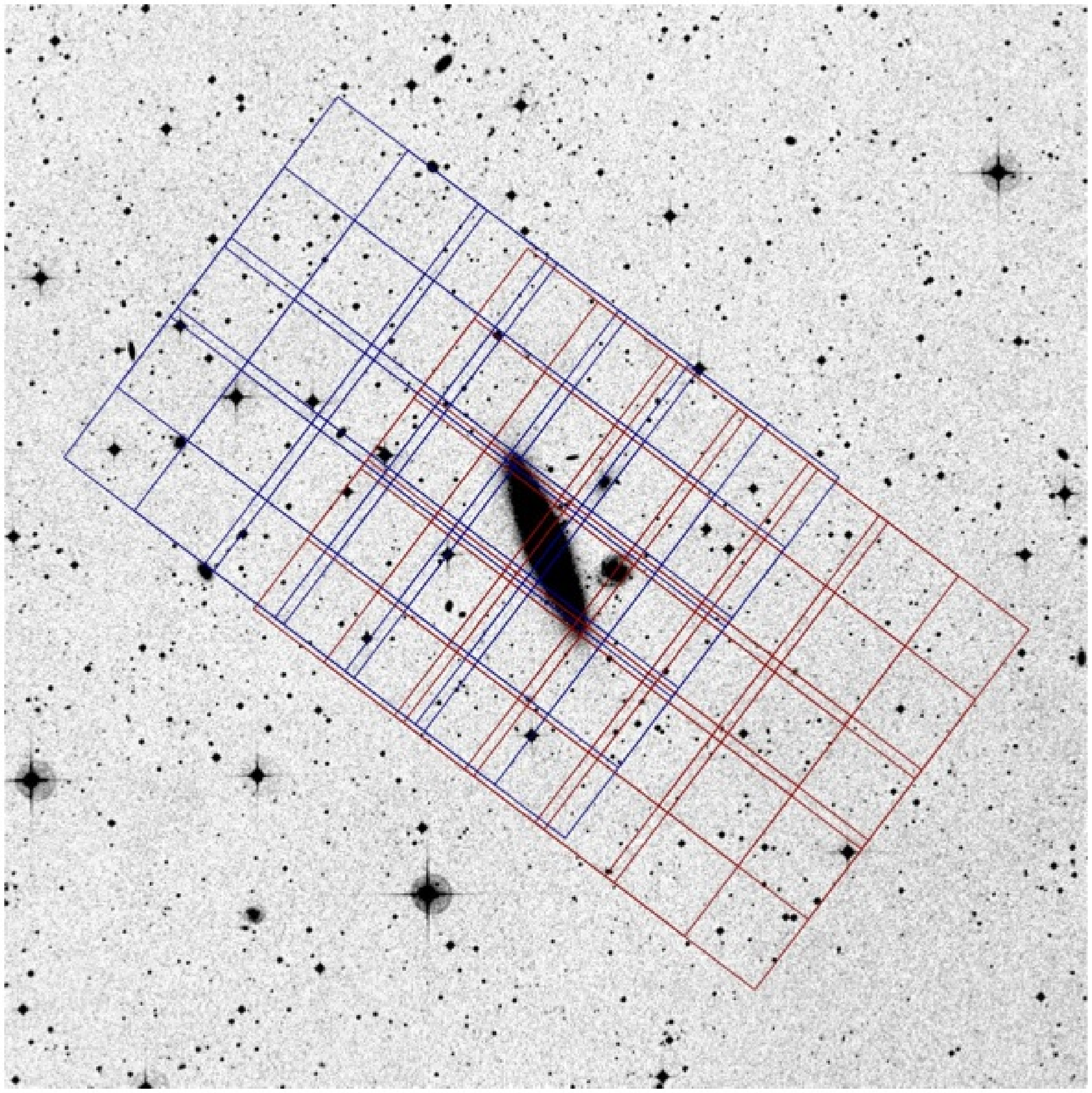}
\caption{ {\em Left:} A typical dithering AOR shown for NGC 2919 in the warm mission.  The blue (red) overlay shows the 3.6$\mu$m (4.5 $\mu$m) coverage.  The galaxy is mapped in two separate AORs separated by at least 30 days.  The flanking fields are reduced by the \sfg pipeline and may also contain serendipitous objects such as interacting galaxies or tidal debris.  {\em Right:} A typical mosaicking AOR shown for NGC 1515 in the warm mission.  The blue grid shows the coverage of the 3.6$\mu$m camera and the red grid shows the coverage of the 4.5 $\mu$m camera. The galaxy is mapped to $>$1.5$\times$D$_{25}$ by both arrays.  The mosaicking pattern takes steps of 146.6$\arcsec$ with 30s integrations in each location.  The total 240 s integration time is obtained in two separate AORs.  Note that significant amounts of sky are also mapped on opposite ends of the galaxy by each of the individual arrays.  } \label{mappingaor}
\end{figure}

1560 of the new observations are of galaxies with D$_{25} < $ 3.3$\arcmin$.  We can map these galaxies to 1.5$\times$D$_{25}$ with a single pointing because 
the {\em Spitzer} field of view is 5$\arcmin$.  Hence  all of these galaxies are
mapped using a standard cycling
small dither pattern with 4 exposures of 30s each in two separate Astronomical Observation Requests (AORs). Each pair of AORs is separated by at least 30 days to allow for sufficient
rotation of the telescope so that the galaxy is imaged at two distinct
orientations.  Two AORs allow us to better remove cosmic rays, and the  
redundant information gathered by the two visits allows us to
characterize and remove image artifacts (e.g., muxbleed, column
pulldown, etc (see the IRAC Instrument Handbook\footnote{ See \tt{http://ssc.spitzer.caltech.edu/irac/iracinstrumenthandbook/}}) and possible asteroids. Drizzling the data from the two visits also allows us to achieve sub-pixel sampling with which we can reconstruct images with better
fidelity than would be possible from a single visit.  We also note
that since both the 3.6 and 4.5$\mu$m arrays collect data
simultaneously, their offset placement in the {\em Spitzer} focal plane
means  that an additional flanking map is made adjacent to the galaxy in
each of the two channels (in some cases, the flanking field from each of the AORs may not overlap exactly due to the telescope rotation, but this does not affect the observations of the main galaxy). An example of a typical dithering AOR is shown in the left panel in Figure \ref{mappingaor}.

One of the considerations in designing these AORs
was the effect of saturation from bright stars on the chip and the
scattered light from bright stars falling in one of the three
scattering zones surrounding the arrays\footnote{A description of this
problem is given in the IRAC instrument handbook at: .http://ssc.spitzer.caltech.edu/irac/iracinstrumenthandbook/}.  Stars
brighter than 5th magnitude can leave a latent charge on the chip.  
Stars brighter than 11th magnitude in the scattering zones may scatter
light into the main field of view of the IRAC chips. To mitigate these
cases, we used the {\em Spitzer} Observing Tool
(SPOT\footnote{\tt {http://ssc.spitzer.caltech.edu/warmmission/propkit/spot/}})
to examine every galaxy with a star, m$_{Ks} <$ 8 (stars fainter than m$_{Ks}$ = 8 do not contribute significantly to the background) within an area of 300$\arcsec$, covering the IRAC field of view and scattering zones adequately, for all possible visibility windows through 2011.      
In a majority of these cases, no modifications of AORs were
needed because the offending star(s) fell out of the chip or the
scattering zones. In the remaining cases, we constructed AORs with
timing constraints, slightly offset pointings or used medium dithers
or mosaics to reduce the effect of the bright stars.
 
For galaxies with D$_{25}$ $>$ 3.3$\arcmin$, we mosaic them in array
coordinates with offsets of 146.6$\arcsec$ with 30s integrations at each
location to create a map $\ge$1.5$\times$D$_{25}$. This leads to a map with each pixel observed four times at all wavelengths in the core of the map. In addition, on each side of the galaxy there
will be a region where each pixel is observed twice but only in one of the two
channels. Like the dithered AORs, each mosaic is observed
twice but with only a follow-on constraint so that each mosaic overlaps the other
closely.  An example of this type of an AOR is shown in the right panel of Figure
\ref{mappingaor}.
 
Although larger mosaics could be made, our experience with SINGS data
showed that the dominant noise term comes from variations of the
background \citep{regan04,regan06}. The best background for a
galaxy includes sufficient sky not far from the area of interest. Our
map sizes are thus fine-tuned to achieve the best possible signal-to-noise for all galaxies.

\section{The \sfg Pipeline} \label{pipeline}

The \sfg pipeline is divided into four distinct parts, which we refer to as Pipelines 1, 2, 3 and 4.  Pipeline 1 takes the raw basic calibrated data (BCD) fits files and converts them to science-ready data.  Pipeline 2 creates a mask for the foreground and background objects  for each of the individual galaxies.  Pipeline 3 measures the standard global galaxy properties such as size, axial ratio, magnitude, color, etc. and computes the radial profiles of the standard properties.  Pipeline 4 deconstructs each galaxy into its major structural components. Each of these is described in detail here.

\subsection{\sfg Pipeline 1}

The purpose of pipeline 1 (P1) is to create science-ready mosaics from the two visits of each target.
To do this P1 performs two major steps.
First, it matches the background level in the individual images to account for drifts in the zero point of the amplifiers.
Second, it creates the final mosaic using the Drizzle package of IRAF\footnote{IRAF is distributed by the National Optical  Astronomy Observatories, which are operated by the Association of Universities for Research in Astronomy, Inc., under cooperative agreement with the National Science Foundation.}.

To match the background levels in the individual images we find the regions of overlap between all pairs of overlapping images.  Typically we require that the overlap region contain 20000 pixels.  For all of the new post-cryogenic mission observations (and archival observations that used the same observing strategies), such an overlap is always available.  However, when reducing other archival data with less overlap between the frames, we reduce the number of pixels in the overlap region to as low as 5000 pixels -- this still provides an adequate number of pixels for estimating the  background.  
 
Within these regions we determine the brightness level of the 20th percentile pixel. We use the 20th percentile brightness level to better avoid detector artifacts such as muxbleed which can affect even the median.
By using the 20th percentile brightness level we have a better chance of using the true sky level.
We then perform a least square solution using the brightness differences between all the pairs of overlapping images, minimizing the residual background difference.
Since we use the difference, we need to use the background level in the first image to set the zero point of the solution. We then make corrected images by adding the solved-for corrections and use these images in the formation of the final mosaic.  We create the final mosaic following the standard prescription of the STSDAS DITHER package.  This method removes the cosmic rays by first forming a mosaic from a median combination of the images and then comparing the individual images to the median mosaic and flagging any cosmic rays.
The final mosaic is formed by drizzling the individual images.  The resulting mosaic has a pixel scale of 0.75$\arcsec$ and we correct for the change in pixel size to keep the units of MJy sr$^{-1}$.  

The relative astrometry of the mosaicked images is excellent because the {\em Spitzer} pipeline already incorporates the 2MASS positions to update its astrometry for each of the basic calibrated data frames.  The \sfg pipeline therefore does not require the type of cross-correlation that was needed previously for the SINGS pipeline to align overlapping fields.  The relative astrometric accuracy is $<$0$\farcs$1.

The point spread function for IRAC is asymmetric and depends on the spacecraft orientation angle.  Pipeline 1 images are produced by combining two different visits to the galaxy and therefore we estimate the PSF by combining stars in the foreground and background for six typical \sfg galaxies to create a "super"-PSF which has a typical FWHM of 1$\farcs$7 and 1$\farcs$6 at 3.6$\mu$m and 4.5$\mu$m respectively.  For the most distant galaxies in our survey at a distance of $\sim$40Mpc, this resolution corresponds to a linear scale of $\sim$300pc. At the median distance of the galaxies in this survey at 21.6 Mpc, this corresponds to a linear scale of $\sim$170pc. 

\subsection{\sfg Pipeline 2}

After the science-ready images are produced by the \sfg pipeline, we generate masks for point sources using SExtractor \citep{bertin96} for both channels.  Each mask is checked by eye and iterated to identify any point source otherwise missed by SExtractor.  We also unmask any region of the galaxy that may be have been incorrectly identified by SExtractor.  
The images with their corresponding masks are then run through the third part of the \sfg pipeline (Pipeline 3, P3 hereafter).  

\subsection{\sfg Pipeline 3}

\begin{figure}

\plotone{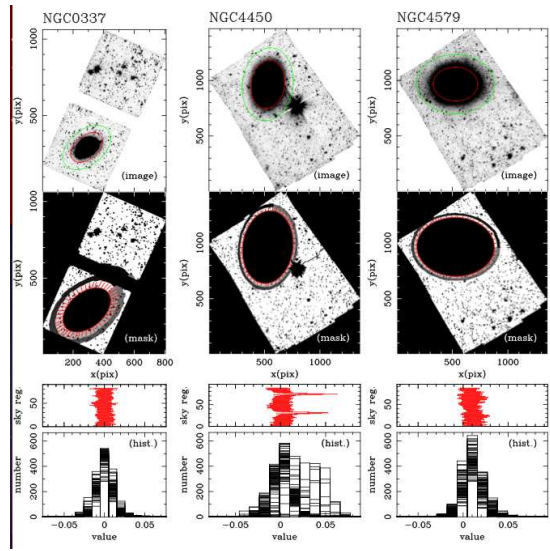}
\caption{Results of the determination of the sky background for the
  \sfg  galaxy NGC~0337, NGC 4450 and NGC4579. Top row: Results for
  the IRAC 3.6$\mu$m image for each galaxy. The D$_{25}$ ellipse is
  shown in red and inner edge of the annulus from where we begin to
  estimate the sky is shown in green.  Second row: The elliptical
  annuli used to compute the sky background are shown.  The red
  segments indicate the 45 regions of the inner annulus.  The outer
  annulus used to trace the gradient is shown shaded by the grey scale
  annulus.  The changing shade from light grey to black corresponds to the 45 regions of the outer annulus.  In this row all the masked pixels are shown in black but note that the black color of the outer annulus does not mean that these regions are masked.  
  Third row: The median and standard deviation for each of the 90 sky
  boxes is shown here.  Bottom row:  A histograms of the pixel values within each of the 90 sky
  boxes is shown in this panel.}\label{ngcsky}
\end{figure}

The first step in P3 is to determine the local sky level around each
galaxy. We want to be sufficiently far from the galaxy to ensure that
there is no contamination from the galaxy but not so far that the
background is not truly local.  We do this by computing the median sky
value in two concentric elliptical annuli centered at the position of
the galaxy. Typically we start the inner annulus at a distance of
1.5$\times$D$_{25}$.  This annulus is divided in azimuth into 45
regions (or sky boxes). Each of these regions is then grown outwards
until it encompasses a total of 4000 non-masked pixels\footnote{For
  archival galaxies that fill the mosaicked frame, this requirement is
  reduced to 1000 non-masked pixels.}, as shown in Figure
\ref{ngcsky}. The second annulus begins at the end of the first
annulus and the same process is repeated.  We compare the measured sky
values in the two annuli for any radial gradients which might be
indicative of light contamination from the galaxy.  If we find that
there is contamination, then we move the inner radius outwards and
repeat the process until we are assured that we are sampling the local
sky.

In our analysis so far, the inner radius of the inner annulus has
varied between 1.2 and 2$\times$D$_{25}$.  In all cases for the warm
mission data and a majority of the archival data, it has been
sufficent to set the inner radius of the annulus to either 1.5$\times$
or 2$\times$ D$_{25}$.  However in a few cases where the archival data
are shallow and the galaxy is relatively blue, we moved the inner
radius to 1.2 $\times$D$_{25}$ to be immediately outside the galaxy to
get as accurate a measurement possible for the local sky level.
Automating this procedure is non-trivial because variations at these
faint levels ($\mu_{3.6\mu m} > $27 (AB) mag arcsec$^{-2}$) can be
from a lopsided galaxy, debris or tidal structures, or variation in
the background.  Therefore we examine every galaxy by eye to ensure
that contamination from the galaxy light to the measurement of the sky
background is as low as possible and move the radii outwards as
needed.

The sky background computation method for three of the galaxies in the
\sfg sample (NGC~0337, NGC~4450, NGC~4579) is shown in Figure
\ref{ngcsky}.  In the central column of this figure, we show an
example case for NGC 4450 where the median sky in some of the boxes is
affected by the bleeding of flux from a (masked) but saturated nearby
bright field star.  In such cases we ignore the affected pixels or
move the annuli further out to get the best estimate for the
background.  Finally, we use the standard deviation obtained from the
90 sky boxes with the average standard deviation within each box to
estimate the contribution of the low-frequency flat-fielding errors
and possible additions to the total error budget in the background
determination.

The typical sky brightness level is fainter than 26
(AB)\,mag\,arcsec$^{-2}$ at 3.6 and 4.5\,$\micron$.  The Poisson noise
from the galaxy dominates the total error budget in the central
regions, while errors in the sky background are the main source of
uncertainty in the outer parts. Large-scale background errors can be
particularly significant in galaxies with large apparent sizes.
However, in the fainter outer parts, the noise is significantly
reduced from azimuthally averaging over a large number of pixels in
the outer galaxy -- we are therefore able to measure the galaxy
profiles well below the typical sky brightness. The typical S/N
  for a \sfg galaxy at the canonical $\mu_{3.6\mu}$ = 26.5 (AB) mag
  arcsec$^{-2}$ is $>$ 3.

The images resulting from Pipeline 1 are calibrated in units of
MJy sr$^{-1}$. The conversion from MJy sr$^{-1}$ to AB magnitudes is such that the
zero-point to convert fluxes in Jy to magnitudes does not vary with
wavelength \citep{oke74}:

\begin{equation}
m_{\mathrm{AB}} (\mathrm{mag})=-2.5\log F_{\nu} (\mathrm{Jy})+8.9
\end{equation}

From this definition one can derive the corresponding expression for surface brightness:

\begin{equation}
\mu_{\mathrm{AB}} (\mathrm{mag})=-2.5\log I_{\nu} (\mathrm{MJy\ str^{-1}})+20.472
\end{equation}

\subsubsection{P3: Measuring the Galaxy Host Properties}

\begin{figure}
\begin{center}
\plotone{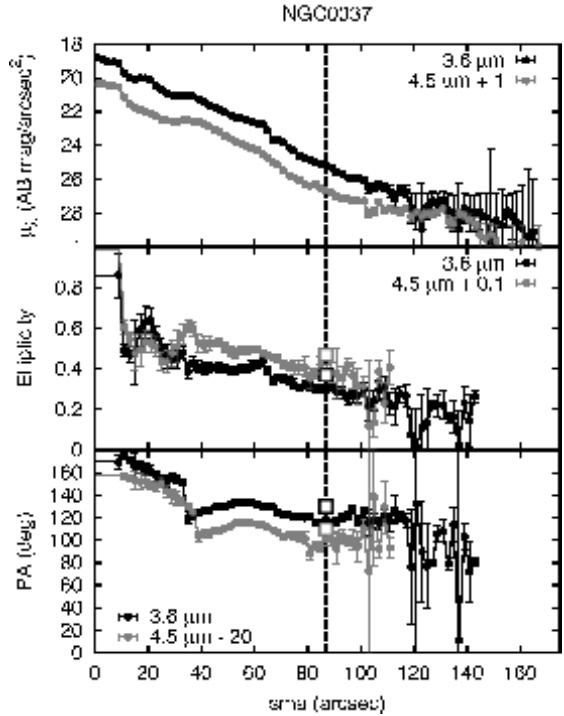}
\caption{Surface photometry results for the \sfg  galaxy NGC~0337. Top: Surface brightness profiles for 3.6$\mu$m (black filled circles) and 4.5$\mu$m+1\,mag (grey filled circles). The vertical dashed line shows the position of the B-band D$_{25}$ major axis radius. Middle: Variation of the ellipticity of the best-fitting isophote with the radius. For the sake of clarity we have applied an offset of 0.1 to the 4.5$\mu$m profile. Bottom: Position Angle of the best-fitting isophote as a function of radius. Note the 20\,dg offset applied to the 4.5$\mu$m profile. The boxes drawn at the position of the D$_{25}$ major axis radius in the latter two panels indicate the ellipticity and PA listed in the RC3 catalog, respectively.}
\label{NGC0337_ellipse}
\end{center}
\end{figure}

\begin{figure}
\begin{center}
\plotone{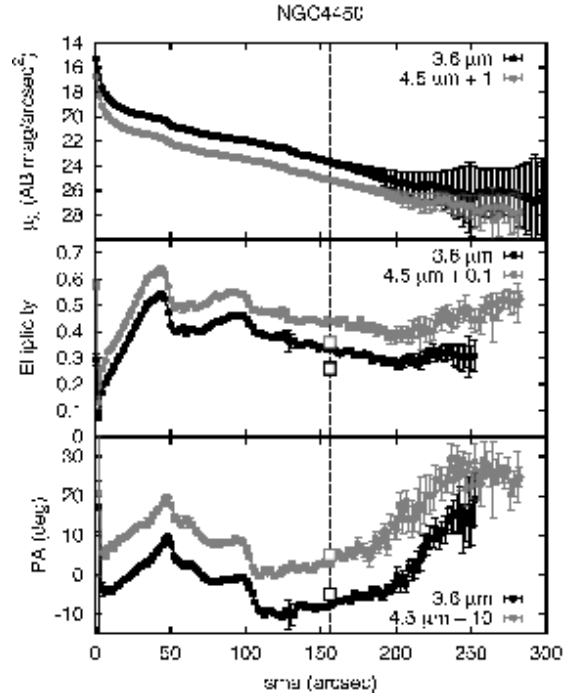}
\caption{The same as Figure~\ref{NGC0337_ellipse}, for NGC~4450.}
\label{NGC4450_ellipse}
\end{center}
\end{figure}

\begin{figure}
\begin{center}
\plotone{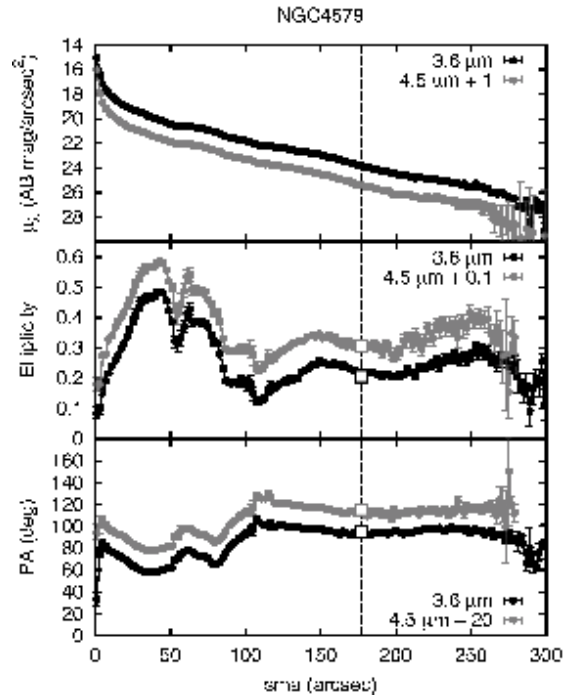}
\caption{The same as Figure~\ref{NGC0337_ellipse}, for NGC~4579.}
\label{NGC4579_ellipse}
\end{center}
\end{figure}

After the sky level has been properly measured, we use the IRAF routine {\tt ellipse} to determine the radial profiles of intensity, surface brightness,  ellipticity and position angle for all galaxies in both bands. The center is determined using {\tt imcenter} and kept fixed during the fit, whereas the ellipticity and position angle are left as free parameters. We generate profiles with two different radial resolutions, by incrementing the semi-major axis in $2\arcsec$ and $6\arcsec$ at each step, respectively. The latter profiles have a coarser resolution and a better S/N ratio - these are used to measure the RC3-like parameters. The low- and high-frequency sky background errors described above,
together with the Poisson noise of the source along each isophote, are
considered together in the final error budget of the surface
photometry at each radius (see \citealt{gildepaz05, munoz-mateos09}). The errors in the position angle and ellipticity at each radius are computed by the ellipse task from the rms of the pixel values along each fitted isophote. The corresponding errors for the values of these geometrical parameters at 25.5 and 26.5 (AB) \,mag\,arcsec$^{-2}$ are interpolated between the corresponding adjacent values.  The error in the semi-major axis is computed from the change in the radius affected by moving the surface brightness profile  by the measured errors and an additional 10\%  error added in quadrature to the measured errors.  The 10\% error is an estimate of the aperture correction uncertainty as estimated by the Spitzer Science Center instrument team.  We compute the error in the semi-major axis for each galaxy.  Sample results for NGC 0337 and NGC 4579 are noted below.  For NGC 4450, the scattered light from a background star strongly affects the computation of the surface brightness error and therefore the error in the semi-major axis that is unrealistic.  In such cases, the error in the semi-major axis can be assumed to be the median semi-major axis error in the \sfg sample.  The maximum error in the determination of the semi-major axis is 6$\arcsec$, the step size of the coarse resolution fitting procedure.
In Figures~\ref{NGC0337_ellipse}, \ref{NGC4450_ellipse} and \ref{NGC4579_ellipse} we show three sample sets of profiles for three galaxies in the sample: NGC~0337 (SBd), NGC~4450 (SAab) and NGC~4579 (SABb). The basic RC3-like galaxy properties (semi-major axis, axial ratio and position angle) are tabulated  at the levels of 25.5 and 26.5\, (AB) \,mag \,arcsec$^{-2}$ in both the 3.6$\mu$m and the 4.5$\mu$m images, and are quoted in Table~\ref{RC3_tab}.

Along with the surface photometry we also measure the curve of growth (e.g., \citealt{munoz-mateos09, gildepaz07}) with the integrated flux up to each radius. By fitting the
accumulated magnitude as a function of the magnitude gradient at each
point we obtain the asymptotic magnitude as the y-intercept of that
fit (see \citealt{gildepaz07}). Once the total magnitude is known, it is straightforward to
locate the radii containing a given percentage of the total galaxy luminosity, from
which concentration indices can be determined. In
Table~\ref{mag_tab} we quote the asymptotic magnitudes of the three
sample galaxies considered here, together with the $C_{31}$ \citep{devaucouleurs77a} and $C_{42}$ \citep{kent85} concentration indices.

\subsection{\sfg Pipeline 4} \label{pipe4}

\begin{figure}
\plotone{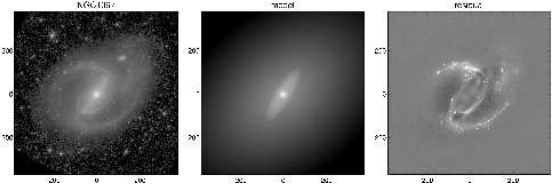}
\caption{Five-component decomposition model for NGC 1097.  The panels show the
observed 3.6 micron image (left), the model image (middle), and the
normalized residual image (OBS-MODEL)/MODEL (right).  North is up and
east is to the left per the usual conventions and the units are in
arcseconds.  The decomposition
includes bulge, disk, bar, central point source and nuclear ring
components, as illustrated in Fig \ref{dc2}. Note how the residual image
emphasizes the outer spiral and the inner ring structures. }\label{dc1}
\end{figure}

\begin{figure}
\plotone{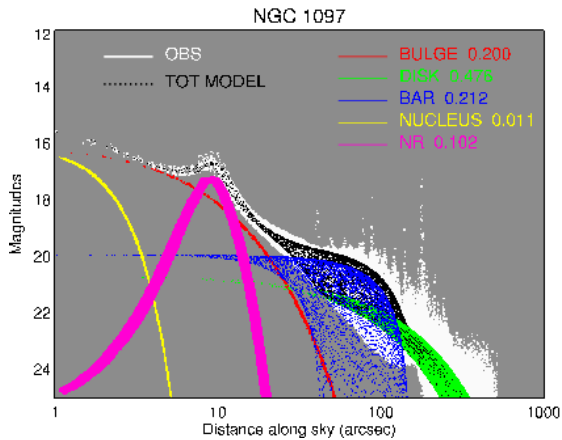}
\caption{The different sub-components of the NGC1097 fit in Fig. \ref{dc1}.  The
white dots show the magnitude of each pixel of the observed image, as
a function of distance from the galaxy center along the sky plane,
while the dark points indicate the total model. The different colors
indicate different sub-components: the numbers after the labels
indicate the relative contribution of the component to the total model
flux. In this type of plot axially symmetric components appear as
curves, while non-axisymmetric components (i.e. the bar) fill a region
confined by their major and minor axis profiles.  Similar plots have
been used to visualize BDBAR \citep{laurikainen05, laurikainen09} and BUDDA
\citep{gadotti08} decompositions.}\label{dc2}
\end{figure}

\begin{figure}
\plotone{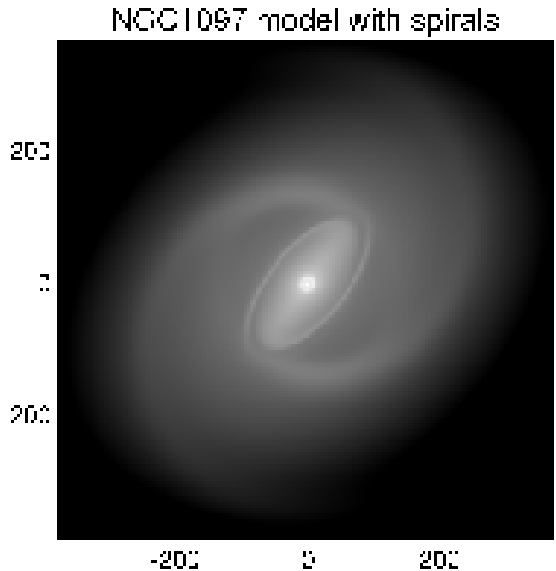}
\caption{Illustration of the capabilities of GALFIT3.0: a decomposition for NGC
1097 including spiral sub-components. While for many scientific purposes it is reasonable to use parametric
functions like S\'ersic or Ferrers functions for fitting
structural components, GALFIT also allows modeling of curved or
irregular structures.  As an illustration of  modeling where the spiral
arms form part of the decomposition, a model for NGC 1097 is shown -- it shows 
a remarkable resemblance to the observed image.  We caution the reader however that since the residuals contain contributions from partially modeled spiral arms, the model parameters output by GALFIT are not necessarily ideal for the study of faint structures in galaxies.}\label{dc3}
\end{figure}

In Pipeline 4, we decompose the two-dimensional stellar distribution in each
galaxy  into different sub-components using the version 3 of GALFIT \citep{peng02, peng10}. GALFIT is a parametric fitting algorithm
that offers a large flexibility on the fitted models; the optimal
solution for the model parameters is found with the
Levenberg-Marquardt algorithm, performing non-linear least-squares
$\chi^2$ minimization of the difference between observed and model
images.  For our pipeline, we could have also used BUDDA \citep{gadotti08} and
BDBAR \citep{laurikainen05}. 
We tested all three algorithms using a set of galaxy
images and verified that GALFIT produces the same results as BUDDA and BDBAR if the three codes are run with the same set of input images, models and initial parameters.  We chose GALFIT  because it is currently the most
sophisticated algorithm, and also well known by a large portion of the astronomical
community. Other studies in which software was developed for similar purposes include \citet{simard98,pignatelli06,mendez-abreu08}.

To simplify the use of GALFIT for a large number of cases,
a set of new IDL-based tools have been created (``GALFIDL''). They include
automatic creation of the input files for decompositions, containing
reasonable first guesses for the initial parameters based on P3
products, as well as routines for running and visualizing the
decompositions.  GALFIDL will be available to the community concurrently with
the publication of the data and the GALFIT analyses; an overview of
these procedures is already available at {\tt http://cc.oulu.fi/$\sim$hsalo/galfidl.html}.

As a standard part of the pipeline analysis, decompositions
include bulges and disks, and, if appropriate, bars,
nuclear components, and in some cases multiple exponential disks.
Taking into account that decompositions, particularly at high
redshift, are often made in a more simple manner, we also compute
the 2-component bulge/disk decompositions, and 1-component S\'ersic
function fits. A decomposition for NGC 1097 is shown in
Figs. \ref{dc1} and \ref{dc2}: the 5-component model includes an
exponential disk, a S\'ersic function for the bulge, and a Ferrers
function for the bar.  The galaxy has an extremely prominent
nuclear ring, which is fit using a Gaussian function, and a nuclear point
source using the corresponding PSF.  Without deep, almost dust-free observations, like those in \sfg, the
measurement of bulge structural parameters, like shape, size and
profile, might in some cases be compromised if the quantification of underlying structures, such as the disks, bars and rings is inaccurate.
GALFIT3.0 allows for even more sophisticated decompositions, which will be
made for selected sub-samples of \sfg galaxies. For example, it is
possible to add more components, or one may want to fit the spiral
arms, of which an example is shown in Fig. \ref{dc3}.

\section{Science Investigations}\label{sfgsci}

In this section we highlight some preliminary results from a number of science investigations that are and will be carried out by the \sfg team.  We also discuss possible studies that will be enabled by the \sfg dataset.

\subsection{The Faint Outskirts of Galaxies}

A major area of study for \sfg is a quantitative analysis of stellar
structures in the faint outer regions of galaxies where the stellar
surface density we trace is extremely low and comparable or lower than
the atomic gas reservoir.  Already, deep optical studies have shown
that, in a majority of spiral galaxies, the outer disks exhibit a
secondary, exponential component.  Stellar disks appear to extend beyond 1.5$\times$ D$_{25}$ (e.g., \citet{pohlen06,   erwin08}) with truncated or anti-truncated profiles. {\it GALEX}
\citep{martin05} has found extended UV (XUV) disks even farther out,
where it was assumed only H\,{\sc i} gas existed (e.g., \citet{gildepaz05,
  thilker05, thilker07, munoz-mateos07, zaritsky07}. Not only is \sfg
significantly deeper than the previous optical data, but it also traces
the stellar profiles into the H\,{\sc i} disk of the galaxy -- in over 800
\sfg galaxies, we image an area $>$3D$_{25}$.

Another observation that has been much debated is the break between
the inner and outer exponentials profiles (e.g., \citealt{bakos08,
  azzollini08}). With \sfg we are measuring the break radius and
comparing it to the disk size and mass. The \sfg catalog of these
breaks will provide strong quantitative constraints for galaxy
evolutionary models.  In particular, this catalog will be key to
quantify the possible role of radial stellar migration in shaping the
outskirts of disks \citep{roskar08, sanchez-blazquez09}.

In elliptical galaxies, deep optical observations have revealed the
presence of faint tidal tail and shell like structures (e.g.,
\citet{bennert08, canalizo07}. \sfg contains 188 ellipticals and
lenticulars -- with this set and the \sfg depth we are quantifying the
frequency and mass of shells and other debris features in the
outskirts of ellipticals and lenticulars.

\subsection{Bulges: Classical, Disk-like, Boxy/Peanut}

Bulges are an inhomogeneous class of objects and different types
have been proposed in the literature \citep{kormendy04,athanassoula05a}.
\citet{kormendy04}
distinguishes bulges from pseudo-bulges, while \citet{athanassoula05a} distinguishes
three categories: classical bulges, boxy/peanut bulges, which are a part of a bar, and
disk-like bulges, formed out of disk material. Several studies have analyzed their
observational properties and compared them with N-body models
\citep[e.g.][]{kuijken95,carollo97,bureau99,erwin03,bureau05,fisher06,drory07,fisher08,gadotti09}. Even
in lenticular galaxies there seems to be evidence for disk-like
bulges, based on structural decompositions (e.g.,
\citealt{laurikainen05, laurikainen07,gadotti08,graham08}) and
kinematics (e.g., \citealt{peletier07, falcon-barroso06}). But until now
few studies have analyzed the structural properties of bulges in a
large number of galaxies; none have done it in a sample as large and
deep as \sfg. Moreover, \sfg encompasses a range of large scale structures,
allowing us to examine the environmental influences on the formation of
bulges. This may have important implications since massive bulges are
expected to form from mergers.

From a preliminary analysis of the \sfg data, \citet{buta10} are
finding that many late type ellipticals in the RC3 have disks, and
thus should instead be classified as S0s, consistent with other deep
near-infrared studies \citep{laurikainen10}. With the \sfg data we
will examine how common the disk structure is in all types of
ellipticals.

Finally, the extremely deep data from \sfg allows us to best determine
the structural properties of the bulge component -- in shallower data,
estimates of shape, size and profile are uncertain when measurements
of underlying components such as disks, bars and rings are
imprecise. As described in \S \ref{pipe4}, the \sfg project is carrying
out detailed decomposition of all the galaxies, and expects to compile
the most detailed data base of bulges of all kinds. The combined
advantages of the \sfg data (deep and insensitive to dust
attenuation) and the uniform reduction pipeline (careful and detailed structural analysis) will allow for a suitable examination of galaxies with composite
bulges, i.e., galaxies hosting more than one bulge category (e.g.,
\citealt{erwin03,barentine10,nowak10}).   The spatial resolution of our survey is adequate for resolving typical bulges (our resolution is $\sim$100 pc at the median distance of the survey volume) but is inadequate for precise measurements of the bulge structure from the inability to resolve compact nuclear structures which are often present at the centers of galaxies.  These unresolved point sources affect the values of Sersic parameters, which on average are overestimated in ground-based optical images compared to HST studies (e.g., \citealt{carollo97, balcells03}).
We also expect that the reduced dust
extinction may reveal triaxial structures in bulges, and thus allow
us to probe beyond the standard symmetric S\'ersic fits.

\subsection{Bars}

Stellar bars are dynamically important structures in galaxies. A
bar-induced gas flow leads to a number of evolutionary effects from
the triggering of circumnuclear starbursts to the build-up of bulges
(\citealt{scoville79, simkin80, zaritsky86, athanassoula92, friedli93, martin94, friedli95,sakamoto99, sheth00, sheth02,sheth05}). The cosmological
redshift evolution of the bar fraction is also an important signpost
of the growth and dynamic maturity of galaxies
(e.g., \citealt{abraham99,sheth03,jogee04, elmegreen04,sheth08}).  
Whereas the previous smaller and shallower studies gave mixed results for the change in the bar fraction with redshift, \citet{sheth08} show that not only does the overall bar fraction in disk galaxies decline significantly over the last  seven
billion years, but also that the decline depends on the galaxy type.  They show that the formation and evolution of bars is strongly correlated with the host galaxy mass, bulge-dominance and presumably their dark matter halo. The \sfg data is critical to establish the precise local frequency of bars and its variation with galaxy type.  
It also provides a measurement of the light fraction in the bar component (e.g.,
\citealt{durbala08,gadotti08,weinzirl09})  and accurate measures of properties such as the bar length, shape and ellipticity (e.g., \citealt{menendez07,erwin05b,gadotti10}).    Several techniques may be used to
derive bar strengths, based on calculating bar torques 
\citep{laurikainen02, buta03}, and 
estimating the bar-spiral arm contrast \citep{elmegreen85}.
Analysis of the relative Fourier intensity profiles of bars is compared with models of
the evolution of barred galaxies (e.g., \citealt{athanassoula02b,
  debattista04}) to constrain the history of their evolution and
interaction with their dark matter halo \citep{athanassoula03}.

Another study of interest is the presence of secondary or nuclear bars
which may be responsible for feeding the central black holes and/or
stellar clusters \citep{shlosman90}.  The \sfg observations,
unaffected by the high and patchy extinction often present in galactic
centers, allow us to detect all nuclear bars with sizes of r $>$ 500
pc (our resolution, 2$\arcsec \sim$ 200 pc at the median distance of
\sfg) in galaxies of all types.

\subsection{Galactic Rings}
 
Outer rings are large, optically low surface brightness features that
dominate the outer disks of some early-type barred and weakly-barred
galaxies (see review by \citet{buta96}. The properties of the rings
constrain their formation epoch, the dynamical time-scale at their
respective radii, and the evolution of the bar pattern speed (e.g.,
\citet{athanassoula03}. Finding outer rings around unbarred galaxies
is also of great interest, because it would suggest that some unbarred
galaxies may in fact be highly evolved former barred galaxies.
Whereas ground-based near-IR imaging is rarely deep enough to detect
outer rings reliably to study their stellar populations, colors, and
other characteristics, the \sfg data will not only allow us to detect
many new outer rings, but also to make a complete census of these
features in normal galaxies and to measure their stellar mass, shape,
width and ellipticity. Likewise, \sfg provides an unprecedented view
of circumnuclear and inner rings, which are typically found in barred
spirals. These rings should have a broad, underlying older stellar
population background as they evolve, and it is this type of feature
we will see in the \sfg data.

\subsection{Spiral Arms}

\begin{figure}
\plotone{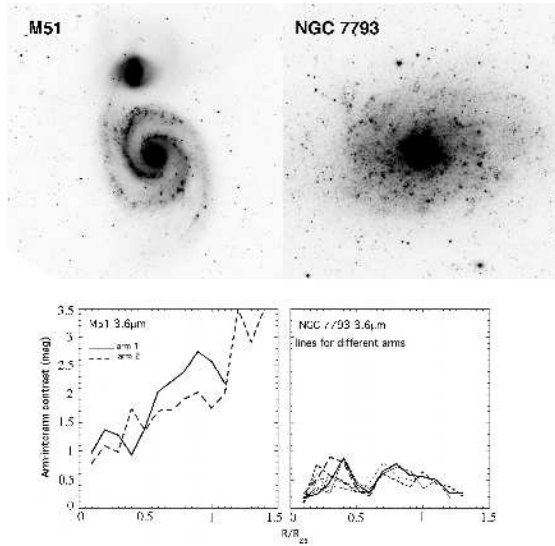}
\caption{Preliminary analysis of the arm-interarm ratios computed for
  a flocculent spiral, NGC~7793, and a grand design spiral, M51, from
  the \sfg 3.6$\mu$m data.  Note the increasing contrast in the grand
  design galaxy compared to the flocculent galaxy.  Detailed analysis
  of the stellar density variations will shed important light on
  spiral density wave theories. }\label{arminterarm}
\end{figure}

\begin{figure}
\plotone{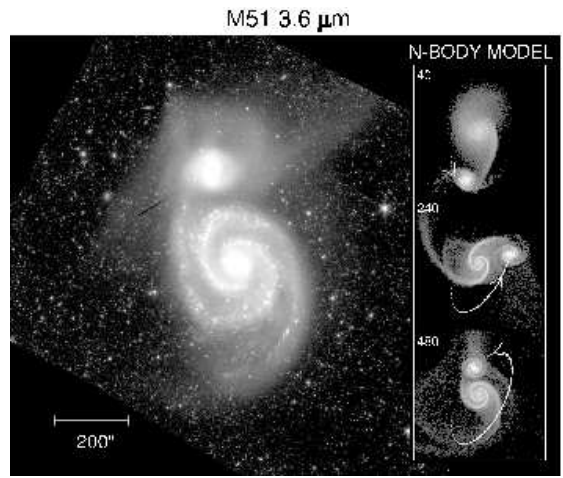}
\caption{Comparison of the M51 3.6 micron image with the N-body simulation model of 
\citet{salo00a,salo00b}. The right hand column shows the simulation at 40, 240 and 480 Myr from the top to the bottom panels respectively.  In the model the near-IR and rotation curve 
data existing at the time were used for setting up the initial disk and halo 
models. The simulation creates an outer southern tail when the companion 
(with mass ratio 0.55) crosses the primary disk towards the observer about
480 Myr ago. To obtain the observed apparent counter-rotating HI kinematics, 
the tail has a tilt of nearly 40 degrees with respect to the inner disk. 
The most recent disk plane crossing about 100 Myr ago is responsible for 
the ejecta north of the companion, drawn from the primary disk, and visible 
in the \sfg image. During the prolonged interaction the tidal wave induced
by the direct perturbation is able to propagate to the central regions, 
consistent with the inner spiral structure described by \citet{zaritsky93}.
}
\label{fig_m51_simu1}
\end{figure}

\begin{figure}
\plotone{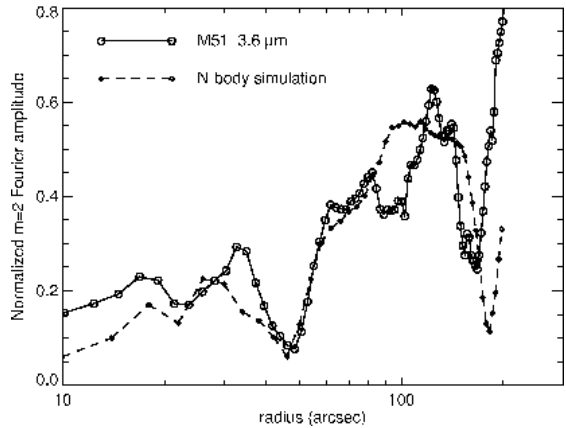}
\caption{Comparison of the m=2 Fourier amplitude profile of
M51 extracted from $S^4G$ $3.6 \mu m$ image with the N-body
simulations of \citet{salo00a,salo00b}. The simulation
corresponds to the final snapshot of Fig. \ref{fig_m51_simu1}. The
outermost observed minimum at $r \approx 180$ arcsec is related to the
change of the curvature of arms: in simulations a similar feature
marks the region where the self-gravity sustained spiral wave
separates from purely tidal outer structure. In the model the minima
at $r \approx 40$ and $20$ arcsec are amplitude modulations, which
arise due to interference between separate inward-propagating tidal
wave packets. See also Figure 11 for M51 at $R_{25} \approx 300''$}
\label{fig_m51_simu2}
\end{figure}

A systematic study of spiral arm amplitudes as a function of radius at
3.6 and 4.5$\mu$m, free of confusing effects from dust extinction,
promises to initiate a new set of spiral arm studies. Although the
discrepancies in the classical density wave model were solved with the
swing amplification theory \citep{toomre81}, there has been very
little testing of this theory with observations. Modal studies by
\citet{bertin89a, bertin89b} predict that there should be amplitude
variations along the arms from resonances between inward and outward
moving spiral waves and the precise amplitude variations have been
modeled \citep{elmegreen84, elmegreen89, elmegreen93,regan97}. Most
recently, a new theory has been proposed for the formation of rings
and spiral arms which argues in favor of chaotic orbits, confined by
invariant manifolds emanating from the L1 \& L2 Lagrangian points of a
bar \citep{romero-gomez07}.  Several comparisons between the 
results of this theory and observations have been carried out successfully
\citep{athanassoula09a, athanassoula09b} and several more will be
possible with the \sfg database.
\sfg will cover the full range of
galaxies with varying spiral arm strengths and galactic structures. In
Figure \ref{arminterarm}, we show a preliminary result of the type of
analysis that is possible with the \sfg data. In this figure we show
the arm-interarm contrast measured from the 3.6$\mu$m \sfg images for
NGC~7793, a flocculent galaxy, in contrast with M51, a grand design
spiral galaxy (see a model for amplitude variation of M51 in \citealt{salo00a}).
The contrasts are measured beyond R$_{25}$, further
than in previous optical studies. The spiral structure traced in the
3.6 $\mu$m band is different in detail from the optical images, but still
shows weak arms in NGC~7793 and increasingly stronger arms in M51.  We
will carry out a detailed analysis of spiral arm amplitudes and
variations and compare the observational data to the models.

For the NGC 5194 (M51) system, there may well be a connection between the spiral  structure and the presence of the companion (NGC 5195). Indeed, detailed HI
observations by \citet{rots90} show the presence of a long tail, which
emanates from the west side of the galaxy, running south of the spiral arm
visible in the \sfg image, and turning up towards the north further eastwards.
The HI kinematics are very difficult to model with an interaction involving 
a single passage, but \citet{salo00a,salo00b} propose a model with a 
second passage (Figure \ref{fig_m51_simu1}), which gives a better fit to some of the aspects of 
the observations. The inner spiral structure in the main spiral is then 
amplified by the action of the companion, and wave interference causes 
amplitude modulations, which are well reproduced in the \sfg observations 
(Figure \ref{fig_m51_simu2}). This shows the interest of the \sfg data for detailed modeling 
of the dynamics of individual galaxies.

\subsection{Early Type Galaxies \& Dwarfs}

\sfg contains 188 galaxies of T-type $<$ 0 and therefore constitutes
one of the largest and most homogeneous mid-infrared data sets for
early type galaxies.  The shapes of ellipticals and lenticulars 
may reflect different evolutionary paths, with wet mergers leading to
disky shapes and dry mergers leading to boxy profiles
\citep{naab06,pasquali07,kang07}. In addition, \citet{kormendy09} shows
that deep imaging can reveal departures from the S\'ersic profiles in elliptical galaxies, which are diagnostics of their formation.  They found a dichotomy in which ellipticals that have cuspy cores at their inner radii can be separated from those which show an excess of light at the center. 
The \sfg data offer us
a unique opportunity to revisit the frequency of the different shapes and profiles 
as a function of stellar mass and environment, free of dust
obscuration effects.

Although dwarfs are the most abundant type of galaxy in the Universe,
their low surface brightnesses of typically $\mu_B >$~22 mag~arcsec$^{-2}$ have restricted detailed observations of these stellar
systems. In particular, the characterization of the underlying stellar
structure in these faint galaxies still remains largely
unexplored. Using an absolute magnitude criterion of $M_B > -17$,
we have identified a preliminary sample of 206 dwarf galaxies in
\sfg. The sample includes 87 late-type DDO dwarfs previously
unobserved with {\em Spitzer}, to complement the 48 which have been
observed. The \sfg dwarfs therefore constitute a representative sample
of nearby dwarf galaxies in which to study the properties of the
underlying stellar component.

Dwarf galaxies in \sfg allow for a detailed analysis of the underlying
stellar disk in both the 3.6 and 4.5$\mu$m data. By deriving the
radial scale lengths for these galaxies, more insight can be gained
into the evolutionary link between dwarfs and their giant
counterparts. At the same time, the relatively large sample of dwarfs
in \sfg gives a unique opportunity to look at how these parameters may
vary with environment. We can also derive stellar masses for the
dwarfs which will be used to assess their stellar mass to dark-matter ratio and will
likely provide crucial evidence to the longstanding debate of whether
these galaxies are truly dark-matter dominated systems (see e.g., \citealt{strigari08}).

In addition to these areas, there are at least three other main areas
of astrophysics that these data will address uniquely: the reported
absence of bulges in some disk galaxies, the vertical stellar
structure, and the Tully-Fisher relationship. There is also likely to
be significant amounts of spin-off science for \sfg with its  
large and accurate inventory of mass and galactic
structure in the nearby Universe, e.g., relationship between the
Tully-Fisher relationship and properties of the fundamental plane
(e.g., \citealt{zaritsky08}), dust heating in outer disks, the
interaction / merger fraction from a study of the debris around these
galaxies, and diagnosis of AGN activity from tracing hot dust very
close to the AGN from the [3.6]-[4.5] color.

\section{Data Release and Team Policy}

The \sfg team is committed to releasing all of the reduced data upon
publication of the entire \sfg datasets from each of the pipelines.
We expect the science-ready images from pipeline 1, the masks from
pipeline 2, the galaxy properties and profiles from pipeline 3 and the
input and output files from Pipeline 4 (including the mask, sigma
images, images without NaN values and the GALFIT input and output
parameter files) to be released on a staggered schedule using NED or
IRSA for long term use by the community.  Currently the products of
each of the pipeline are expected to be released in the supplement
issue of the journals upon verification of their quality which will be
on-going and occur as part of the various scientific investigations
conducted by the team.  We expect that some of the data and pipeline
may need to be refined or enhanced in the future.  As the data are
still being acquired (data acquisition is expected to be complete in
2011) and analyzed, our earliest release date for the the
science-ready images (P1 products) for the entire \sfg data is not
expected at least until 2012, and the supplement papers for the
products from the other pipelines are expected to follow within a year
after the Pipeline 1 supplement paper.  In the meanwhile, to allow for
maximum use of the data for science by the larger community, the
collaboration has agreed on an open-door policy whereby any member of
the astronomical community may join the team temporarily as a guest
and pursue any type of science.  The use of the data, publication
policies and authorship on papers are clearly defined in our policy
statement available here: {\tt
  http://www.cv.nrao.edu/$\sim$ksheth/
s4g/S4G\_policy\_v6.pdf}.  This
statement is adopted by the \sfg team and all of its guests.
Interested members of the astronomical community are invited to read
the policy and contact the PI or an \sfg member to discuss the
possibilities of joining the team.

The noteworthy point of this policy is that our team has agreed to avoid carving out
large science categories or areas to encourage the maximum use of the
data within the team, and within the larger community, for any science
that may be possible with these data.  Our philosophy is that if
multiple people are interested in a topic but want to work in separate
teams then we will encourage both teams to collaborate and address the
scientific problems in independent but parallel investigations. 

\section{Summary}

\sfg will provide the deepest, largest, and most homogenous data set of nearby galaxies at 3.6 and 4.5$\mu$m. The spatial resolution will be unmatched at these wavelengths, $\sim$2$\arcsec$, which is ideal for comparisons to ground based optical and near-IR observations, as well as MeerKAT, eVLA, ALMA and other future large radio array surveys.  The sample of 2,331 galaxies includes most morphological types and masses. The images will give an unparalleled view of stellar mass distributions and faint peripheral structures.  All of the galaxies, as well as many additional galaxies from the {\em Spitzer} archives, will be reduced in the same way, giving data products that span the full range from science-ready images to measurements of global galaxy properties and individual components. \sfg is designed to be a useful reference for many years to come. It will contain the best available mid-infrared data for individual galaxies and be among the most complete surveys for statistical studies. It will be useful to study the origin and evolution of galaxies and their dynamical components, and to supplement observations of nearby galaxies at other wavelengths. 

\section{Acknowledgements}
\acknowledgments
The authors thank the referee for their useful comments and suggestions that greatly helped improve this paper.  We are also grateful to the dedicated staff at the Spitzer Science Center for their support and help with the planning and execution of this legacy exploration program.  KS would like to thank L. Armus, E. Bell, S. Carey, E. Churchwell, M. Dickinson,  G. Helou,  N. Scoville, and J. Stauffer for sharing their experiences in leading large teams.   A.G.dP and J.C.M.M are partially financed by the Spanish Programa Nacional de Astronom\'ia y Astrof\'isica under grants AyA2006-02358 and AyA2009-10368. A.G.dP is also financed by the Spanish Ram\'on y Cajal program. J.C.M.M. acknowledges the receipt of a Formaci\'on del Profesorado Universitario fellowship.  EL and HS acknowledge support from the Academy of Finland. KMD is supported by an NSF Astronomy and Astrophysics Postdoctoral Fellowship under award AST-0802399.  DME acknowledges support from the Spitzer Science Center from NASA grant JPL RSA-1368024.  RB acknowledges support from NSF grant AST 05-07140.  EA and AB thank the Centre National d'Etudes Spatiales and ANR-06-BLAN-0172 for support.  K.L.M. acknowledges funding from the Peter and Patricia Gruber Foundation as the 2008 IAU Fellow, from the University of Portsmouth, and from SEPnet (www.sepnet.ac.uk). 
This work is based on observations and archival data obtained with the Spitzer Space Telescope, which is operated by the Jet Propulsion Laboratory, California Institute of Technology under a contract with NASA. Support for this work was provided by NASA.  KS and other staff at the NRAO acknowledge support from the National Radio Astronomy Observatory, which is a facility of the National Science Foundation operated under cooperative agreement by Associated Universities, Inc.

{\it Facilities:} \facility{{\em Spitzer} Space Telescope}

\bibliography{mymasterbib}

\begin{deluxetable}{rrrrrrr}
\tabletypesize{\normalsize}
\tablecolumns{6}
\tablecaption{The \sfg Sample.\label{s4gsamptab}}
\tablewidth{0pt}
\tablehead{
\colhead{Galaxy} & \colhead{RA (J2000)} & \colhead{DEC (J2000)} &
\colhead{T} & \colhead{M$_B$} & \colhead{Log(D$_{25}$)} & \colhead{m$_{Bcorr}$}}
\startdata
UGC12893  &  0.00784  &  17.21952  &  8.4  &  -15.97  &  1.06  &  15.14  \\ 
PGC000143  &  0.03292  &  -15.46140  &  9.9  &  -15.83  &  2.04  &  10.00  \\ 
ESO012-014  &  0.04511  &  -80.34825  &  9.0  &  -18.17  &  1.18  &  13.72  \\ 
NGC7814  &  0.05418  &  16.14554  &  2.0  &  -20.04  &  1.64  &  10.96  \\ 
UGC00017  &  0.06198  &  15.21822  &  9.0  &  -16.02  &  1.39  &  14.60  \\ 
NGC7817  &  0.06635  &  20.75182  &  4.1  &  -21.11  &  1.52  &  11.56  \\ 
ESO409-015  &  0.09223  &  -28.09915  &  6.2  &  -15.16  &  1.02  &  14.43  \\ 
ESO293-034  &  0.10559  &  -41.49592  &  6.0  &  -18.68  &  1.40  &  12.67  \\ 
NGC0007  &  0.13916  &  -29.91678  &  4.9  &  -18.22  &  1.39  &  13.18  \\ 
NGC0014  &  0.14618  &  15.81659  &  9.8  &  -18.25  &  1.18  &  12.35  \\ 
IC1532  &  0.16464  &  -64.37161  &  4.0  &  -17.89  &  1.26  &  13.86  \\ 
... & ... & ... & ... & ... & ... & ... \\ 
... & ... & ... & ... & ... & ... & ... \\ 
NGC7798  &  23.99042  &  20.74986  &  4.1  &  -20.20  &  1.20  &  12.56  \\ 
NGC7800  &  23.99354  &  14.80723  &  9.7  &  -19.45  &  1.24  &  12.61  \\ 
\enddata
\tablecomments{ All properties obtained from the HyperLEDA database
(http://leda.univ-lyon1.fr/) \citep{paturel03}}

\tablecomments{RA and DEC are decimal hours and degrees in J2000; T is
  the morphological type code (-5:E, -3:E-S0, 0: S0a, 3:Sb, 7:Sd,
  10:Irr); M$_B$ is the absolute B-magnitude; and ) Log D$_{25}$ is
  the apparent major axis diameter at 25 mag arcsec $^{-2}$. D25 is in
  units of 0.1 arcminutes; m$_{Bcorr}$ is the total apparent corrected
  B-magnitude, corrected for inclination, galactic extinction and
  K-correction. A value of -99 indicates that the the value was
  missing from the HyperLEDA database}

\tablecomments{The full table is available in the electronic version
  of the journal.} 
\end{deluxetable}

\begin{deluxetable}{lrrrrrrrr}
\tabletypesize{\normalsize}
\tablecolumns{9}
\tablecaption{Sample RC3-like parameters\label{RC3_tab}}
\tablewidth{0pt}
\tablehead{
\colhead{} & \colhead{} & \multicolumn{3}{c}{3.6$\mu$m} & & \multicolumn{3}{c}{4.5$\mu$m}\\
\cline{3-5} \cline{7-9}
\colhead{Galaxy} & \colhead{$\mu$} & \colhead{sma} & \colhead{ellip.} & \colhead{PA} & & \colhead{sma} & \colhead{ellip.} & \colhead{PA}}
\startdata
\multirow{2}{*}{NGC~0337} & 25.5 & 90.6$\pm$1.5 & 0.287$\pm$0.018 & 118.4$\pm$2.0 & & 84.1$\pm$1.3 & 0.278$\pm$0.037 & 118.5$\pm$4.5\\
& 26.5 & 110.4$\pm$4.7 & 0.270$\pm$0.027 & 119.8$\pm$3.2 & & 106.9$\pm$1.3 & 0.287$\pm$0.055 & 123.8$\pm$7.1\\
\hline
\multirow{2}{*}{NGC~4450} & 25.5 & 209.0$\pm$6 & 0.293$\pm$0.016 & 1.4$\pm$1.8 & & 199.9$\pm$... & 0.293$\pm$0.020 & 3.2$\pm$2.3\\
& 26.5 & 253.0$\pm$6 & 0.336$\pm$0.048 & 19.8$\pm$5.1 & & 266.7$\pm$...  & 0.418$\pm$0.026 & 18.1$\pm$2.3\\
\hline
\multirow{2}{*}{NGC~4579} & 25.5 & 256.0$\pm$2.6 & 0.295$\pm$0.011 & 95.2$\pm$1.2 & & 229.0$\pm$2.6 & 0.268$\pm$0.018 & 95.5$\pm$2.2\\
& 26.5 & 278.6$\pm$17 & 0.251$\pm$0.018 & 78.6$\pm$2.4 & & 263.4$\pm$12 & 0.282$\pm$0.027 & 98.3$\pm$3.4\\
\enddata
\tablecomments{RC3-like parameters for three sample galaxies. For each galaxy and band we quote the semi-major axis (in arcseconds), the ellipticity and the position angle (in degrees) at two levels of surface brightness: 25.5 and 26.5\,(AB)\,mag\,arcsec$^{-2}$.  The error in the semi-major axis is measured by noting the change in the corresponding radius when the surface brightness profiles are shifted up and down by the local measurement error plus 10\% additional error, which is the maximum expected from any scattered light on the detector.  For NGC 4450, the error cannot be measured due to contamination from a background star and is therefore set to the maximum error, 6$\arcsec$, as discussed in the text.  The errors in the ellipticity and position angle are from the ellipse fitting and reflect the rms in the formal fit of the corresponding isophote.  These errors are significantly larger than any error introduced by the shifting the profiles to measure the error in semi-major axis.}
\end{deluxetable}

\begin{deluxetable}{lrrrrrrr}
\tabletypesize{\normalsize}
\tablecolumns{8}
\tablecaption{Magnitudes and Concentration Indices for \sfg Galaxies \label{mag_tab}}
\tablewidth{0pt}
\tablehead{
\colhead{} & \multicolumn{3}{c}{3.6$\mu$m} & & \multicolumn{3}{c}{4.5$\mu$m}\\
\cline{2-4} \cline{6-8}
\colhead{Galaxy} & \colhead{mag (AB)} & \colhead{$C_{31}$} & \colhead{$C_{42}$} & & \colhead{mag (AB)} & \colhead{$C_{31}$} & \colhead{$C_{42}$}}
\startdata
NGC~0337 & 11.432$\pm$0.001 & 2.67 & 2.74 & & 11.888$\pm$0.002 & 2.82 & 2.90\\
NGC~4450 & 9.630$\pm$0.001 & 4.58 & 4.17 & & 10.112$\pm$0.001 & 4.56 & 4.16\\
NGC4579 & 9.083$\pm$0.001 & 5.29 & 4.54 & & 9.564$\pm$0.001 & 5.60 & 4.71\\
\enddata
\end{deluxetable}

\end{document}